\documentclass[12pt]{article}
\usepackage{graphicx}

\newcommand\pubnumber{}
\newcommand\pubdate{\today}

\def\napoli{Department of Physics\\
Carnegie Mellon University, Pittsburgh, PA, 15213}
\def\support{\footnote{Work supported by the Department of Energy, 
          United States, under contract DE-FG02-91ER40682.}}

\def\Title#1{\begin{center} {\Large #1 } \end{center}}
\def\Author#1{\begin{center}{ \sc #1} \end{center}}
\def\Address#1{\begin{center}{ \it #1} \end{center}}

\newcommand\pubblock{\rightline{\begin{tabular}{l} \pubnumber\\
         \pubdate  \end{tabular}}}
\newenvironment{Abstract}{\begin{quotation}  }{\end{quotation}}
\newenvironment{Presented}{\begin{quotation} \begin{center} 
             PRESENTED AT\end{center}\bigskip 
      \begin{center}\begin{large}}{\end{large}\end{center} \end{quotation}}
\def\Acknowledgements{\bigskip  \bigskip \begin{center} \begin{large}
             \bf ACKNOWLEDGEMENTS \end{large}\end{center}}

\def\beq{\begin{equation}}
\def\eeq#1{\label{#1}\end{equation}}
\def\eeqn{\end{equation}}

\def\beqa{\begin{eqnarray}}
\def\eeqa#1{\label{#1}\end{eqnarray}}
\def\eeqan{\end{eqnarray}}

\let\bar=\overbar

\def\Dslash{\not{\hbox{\kern-4pt $D$}}}
\def\dslash{\not{\hbox{\kern-2pt $\del$}}}

\def\msb{{\bar{\ssstyle M \kern -1pt S}}}

\begin{document}
\begin{titlepage}
\pubblock

\vfill
\Title{Review of D Semi-leptonic Decays}
\vfill
\Author{ Chunlei Liu\support}
\Address{\napoli}
\vfill
\begin{Abstract}
Semi-leptonic $D$ decays continue to play an important role in the field of flavor physics. During this presentation, recent measurements
from pseudo-scalar to pseudo-scalar modes, pseudo-scalar to vector modes, and rare modes will be discussed. These results are important for many purposes, such as validating the machinery of lattice QCD, extracting CKM matrix elements, 
and searching for new physics and new interactions.

\end{Abstract}
\vfill
\begin{Presented}
The 5th International Workshop on Charm Physics\\
Honolulu, Hawai'i ,  May 14--17, 2012
\end{Presented}
\vfill
\end{titlepage}
\def\thefootnote{\fnsymbol{footnote}}
\setcounter{footnote}{0}
%




\section{\boldmath Semileptonic $D$ Decays in the Big Picture}

\begin{figure}[htb]
\centering
\includegraphics[height=2.5in,width=0.6\textwidth]{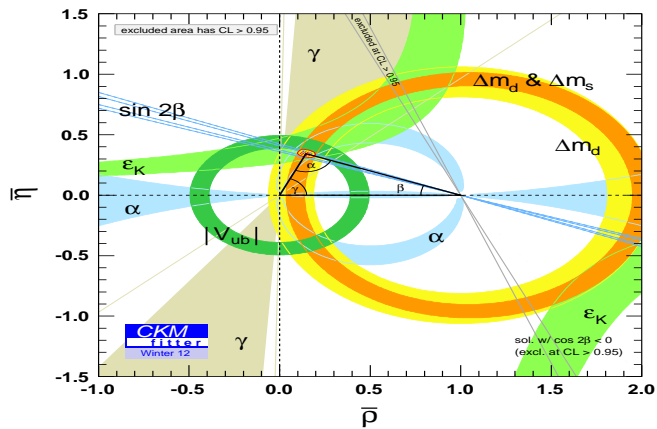}
\caption{ 
Constraints in the $(\bar{\rho}, \bar{\eta})$ plane. The red hashed
region of the global combination corresponds to 68\% C.L..
}
\label{fig:ckm}
\end{figure}

One important task in the field of flavor physics is to over-constrain the CKM matrix. By doing so not only
will we understand better the standard model (SM) physics, but we also have the opportunity to discover 
new physics. To best extract CKM matrix elements, we need inputs both from 
experimental data and from theoretical calculations. Fig.~\ref{fig:ckm} shows our best knowledge about the CKM matrix up to today\cite{ckm}. Our understanding
about sin$2\beta$ is the best thanks to clean theory and large statistics from experiments.  In contrast, our understanding of both 
mixing and $|V_{ub}|$ have limitations from theoretical predictions. 

For example, $|V_{ub}|$ is measured from
$B^0\rightarrow \pi e \nu$ which has the differential decay rate as the following:
\begin{equation}
\frac{d\Gamma}{dq^2} \;=\;  \frac{G^2_F}{24\pi^3}|V_{ub}|^2 p^3_{\pi} |f_+(q^2)|^2.
\end{equation}
In order to extract the value of $|V_{ub}|$ from the branching fraction measurement, we need information on the hadronic form factor $f_+(q^2)$. 
The latest world average of $|V_{ub}|$ is~\cite{ckm}:
\begin{equation}
|V_{ub}|\times 10^3 = 3.92 \pm  0.09 \pm 0.45, \nonumber
\end{equation}
where the uncertainties are from experiment and theory respectively. As we can see, the error due to theory is much larger than the experimental one.
Our best understanding about $f_+(q^2)$ is from lattice QCD calculations. 
To validate improved lattice QCD calculations for the 
form factor in semi-leptonic $B$ decays, we can use semi-leptonic $D$ decays 
as a test and calibration.
The differential decay rate function for the semi-leptonic $D$ decays is :
\begin{equation}
\frac{d\Gamma}{dq^2} \;=\;  \frac{G^2_F}{24\pi^3}|V_{cs(d)}|^2 p^3_{K(\pi)} |f_+(q^2)|^2
\end{equation}
where a massless lepton is assumed. Since the CKM matrix elements $|V_{cs(d)}|$ can be obtained precisely from
unitarity, we have a reliable method to check the lattice QCD calculations.  

The above example demonstrates how charm physics, in particular semi-leptonic $D$ decays, can contribute to the larger picture of flavor physics.  
And more specifically, there are several advantages in using these decays:
\begin{itemize}
\item Key modes have large branching fractions and are experimentally 
easily accessible at threshold experiments (where the missing neutrino 
can be inferred from missing four-momentum).  
\item The theory of semi-leptonic decays is relatively clean, 
since the weak and strong interaction portions can  be factored apart. 
\item The pseudo-scalar to pseudo-scalar decay modes 
provide a simple and clean way to measure the form factor. 
And pseudo-scalar to vector decay modes provide access to 
even more form factors, if desired. 
\item Rare semi-leptonic decays can also provide good paths 
to find new physics or new interactions.

\end{itemize}

\section{Recent Results}

The experimental results of $D$ semi-leptonic decays can be summarized in two main categories: the exclusive decays and the inclusive decays. 
Several recent measurements from the exclusive semi-leptonic $D$ decays are discussed here.

\subsection{\boldmath Measurements of $D \rightarrow K e \nu$ and $D \rightarrow \pi e \nu$}

As previously discussed, these channels provide constraints on lattice QCD given CKM matrix elements or vice versa, and they have
been measured in many experiments such as FOCUS, Belle, BaBar and CLEO-c. 
The main goal currently is to improve the precision 
for the Cabibbo-suppressed decay $D \rightarrow \pi e \nu$. 

The BESIII experiment has taken $\sim$2.9 fb$^{-1}$ $\psi(3770)$ data during 
the 2010 and 2011 data runs. Using one-third of the data, a partially-blind 
analysis has been done with the $D^0\rightarrow K e \nu$ and $D^0 \rightarrow \pi e \nu$ decays.  Preliminary results were first presented at this 
Charm 2012 conference. 

\subsubsection{ Event Selection}
\begin{figure}[htb]
\centering
\includegraphics[height=1.5in,width=0.45\textwidth]{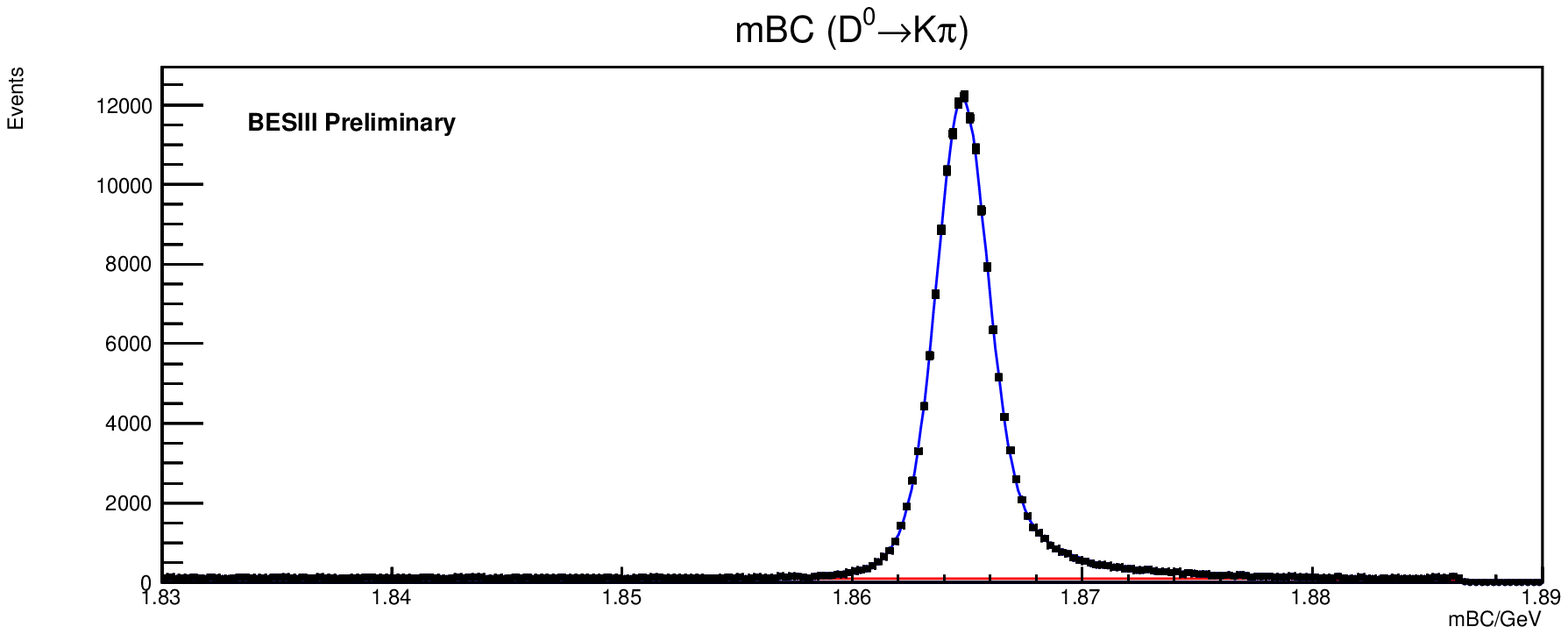}
\includegraphics[height=1.5in,width=0.45\textwidth]{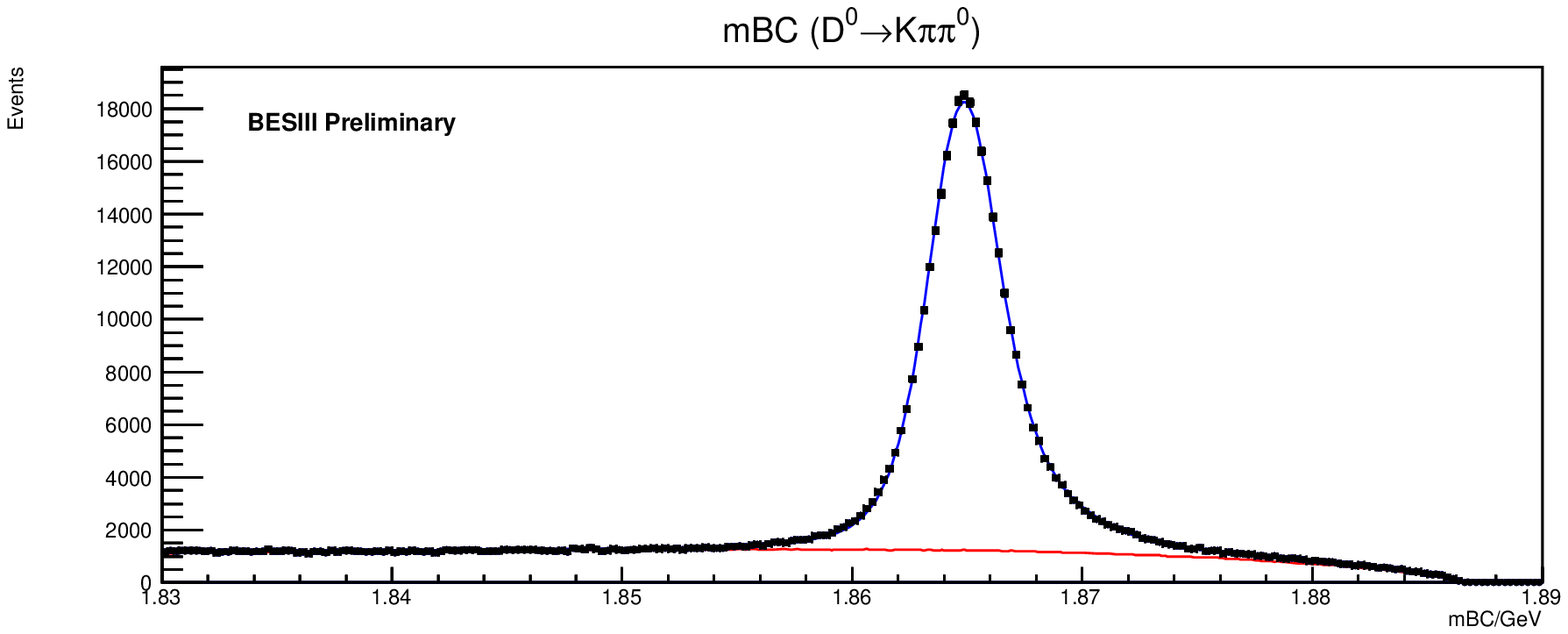}
\includegraphics[height=1.5in,width=0.45\textwidth]{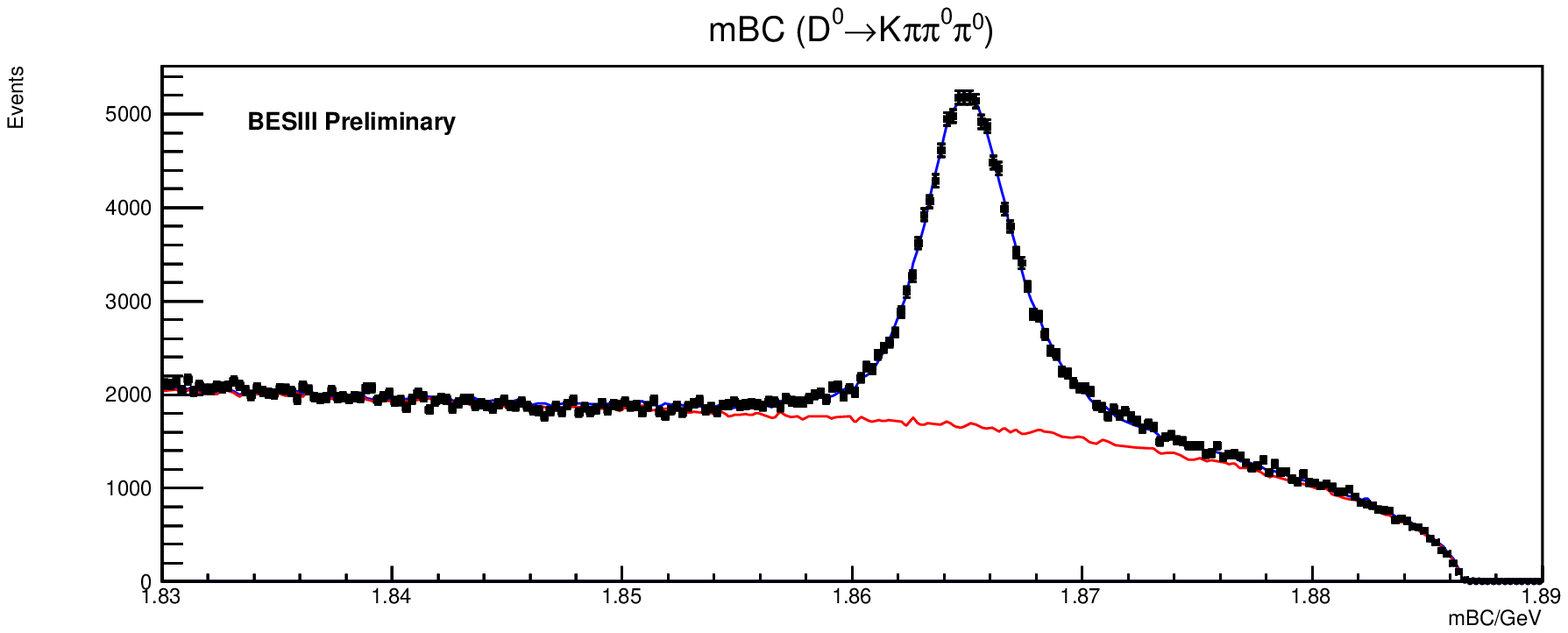}
\includegraphics[height=1.5in,width=0.45\textwidth]{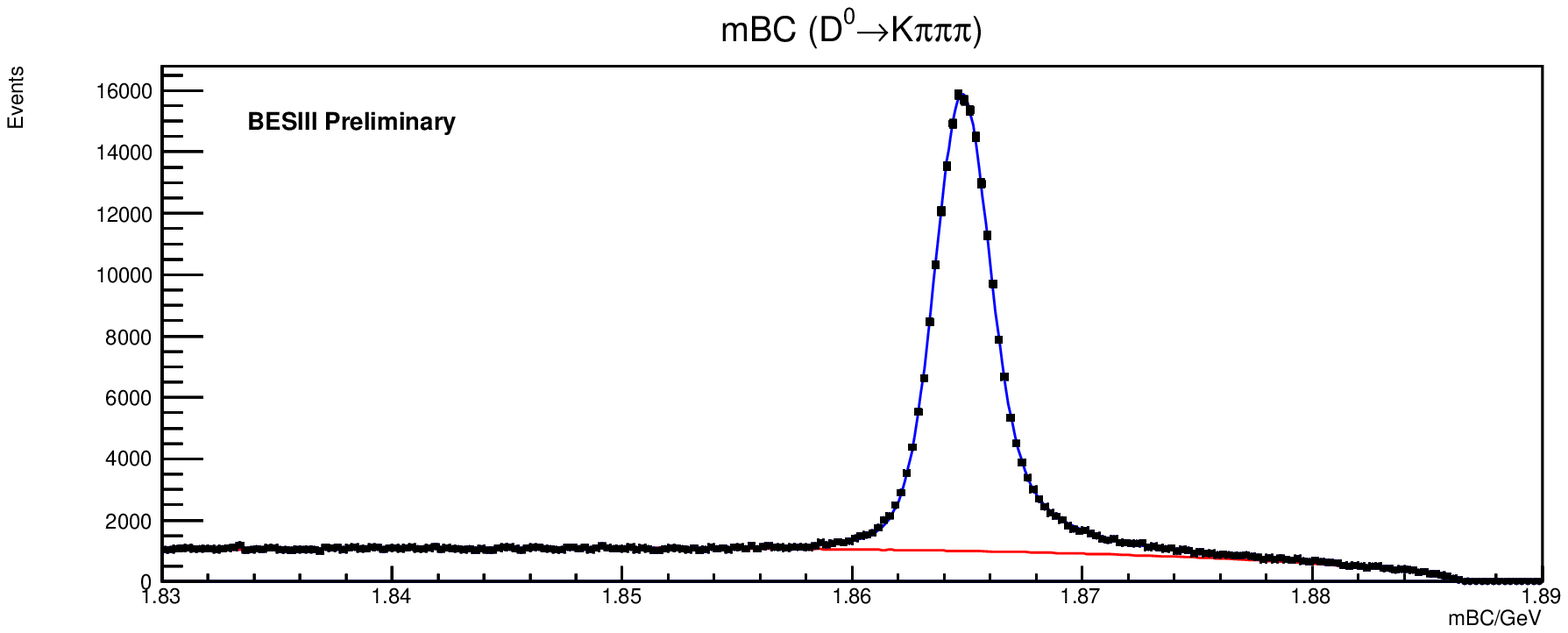}
\caption{ $m_{BC}$ distributions of  $D^0\rightarrow K^-\pi^+$, 
$D^0\rightarrow K^-\pi^+\pi^0$, $D^0\rightarrow K^-\pi^+\pi^0\pi^0$, 
$D^0\rightarrow K^-\pi^+\pi^-\pi^+$, with a $\Delta E$ cut of  $\pm 3\sigma$ for modes without $\pi^0$ and $-4\sigma$ to $+3\sigma$ for modes with $\pi^0$. }
\label{fig:tag}
\end{figure}

The BESIII experiment takes data at the BEPCII symmetric electron-positron collider, with the $\psi(3770)$ produced at threshold.  
Using the double tag technique, several hadronic $D$ decays are fully 
reconstructed at first. 
In case of multiple candidates, the candidate with minimum $\Delta E$ 
is chosen; $\Delta E$ is defined as 
\begin{equation}
\Delta E \;\equiv\;E_{cand} - E_{beam},
\end{equation}
where $E_{cand}$ is the energy of the reconstructed tag mode, and $E_{beam}$ is the beam energy.
The following four hadronic $D$ decays are used: $D^0\rightarrow K^-\pi^+$, $D^0\rightarrow K^-\pi^+\pi^0$, $D^0\rightarrow K^-\pi^+\pi^0\pi^0$,
$D^0\rightarrow K^-\pi^+\pi^-\pi^+$. The $m_{BC}$ distributions of these hadronic tags with our nominal $\Delta E$ requirements are shown in Fig.~\ref{fig:tag}, 
where the beam constrained energy $m_{BC}$ is defined as:
\begin{equation}
m_{BC} \;\equiv\; \sqrt{ E^2_{beam}-|\vec{p}|^2_{cand}},
\end{equation}
where $\vec{p}_{cand}$ is the momentum of the reconstructed tag mode.

\begin{table}[htb]
\begin{center}
\begin{tabular}{l|ccc}  
Tag mode &  Tag Yields &  Fraction(\%) & Tag efficiency \\ \hline
 $D^0\rightarrow K^-\pi^+$ &   159929$\pm$413     &     20.7     &    62.08$\pm$0.07  \\
 $D^0\rightarrow K^-\pi^+\pi^0$ &   323348$\pm$667     &    41.8      &    33.56$\pm$0.03 \\ 
 $D^0\rightarrow K^-\pi^+\pi^0\pi^0$&    78467$\pm$480     &     10.1    &      14.93$\pm$0.04  \\
 $D^0\rightarrow K^-\pi^+\pi^-\pi^+$&    211910$\pm$550     &     27.4     &     36.80$\pm$0.04  \\
\hline
\end{tabular}
\caption{Tag yields and tag efficiencies using $\sim923$ pb$^{-1}$ of $\psi(3770)$ data from BESIII.}
\label{tab:tag}
\end{center}
\end{table}

After hadronic $D$ tags are found, $m_{BC}$ are further required to be between 1.858 GeV and 1.874 GeV before attempting to reconstruct signal candidates.
The corresponding tag yields and tag efficiencies are listed in Tab.~\ref{tab:tag}.
To search for the signal candidates, the following requirements are applied:
\begin{itemize}
\item Only two good charged tracks with opposite charges left in the event.
\item One track is identified as an electron candidate, 
 and the other track is identified as a kaon (pion) candidate for 
 the $K^+ e^- \nu$ ($\pi^+ e^- \nu$) mode. 
\item The electron has the same charge as the kaon track from the tag side $D$.
\item The candidate is vetoed if the most energetic unmatched shower has energy greater than 250 MeV, in order to suppress backgrounds with extra $\pi^0$s.
\end{itemize}
 
\begin{figure}[htb]
\centering
\includegraphics[height=1.5in,width=0.45\textwidth]{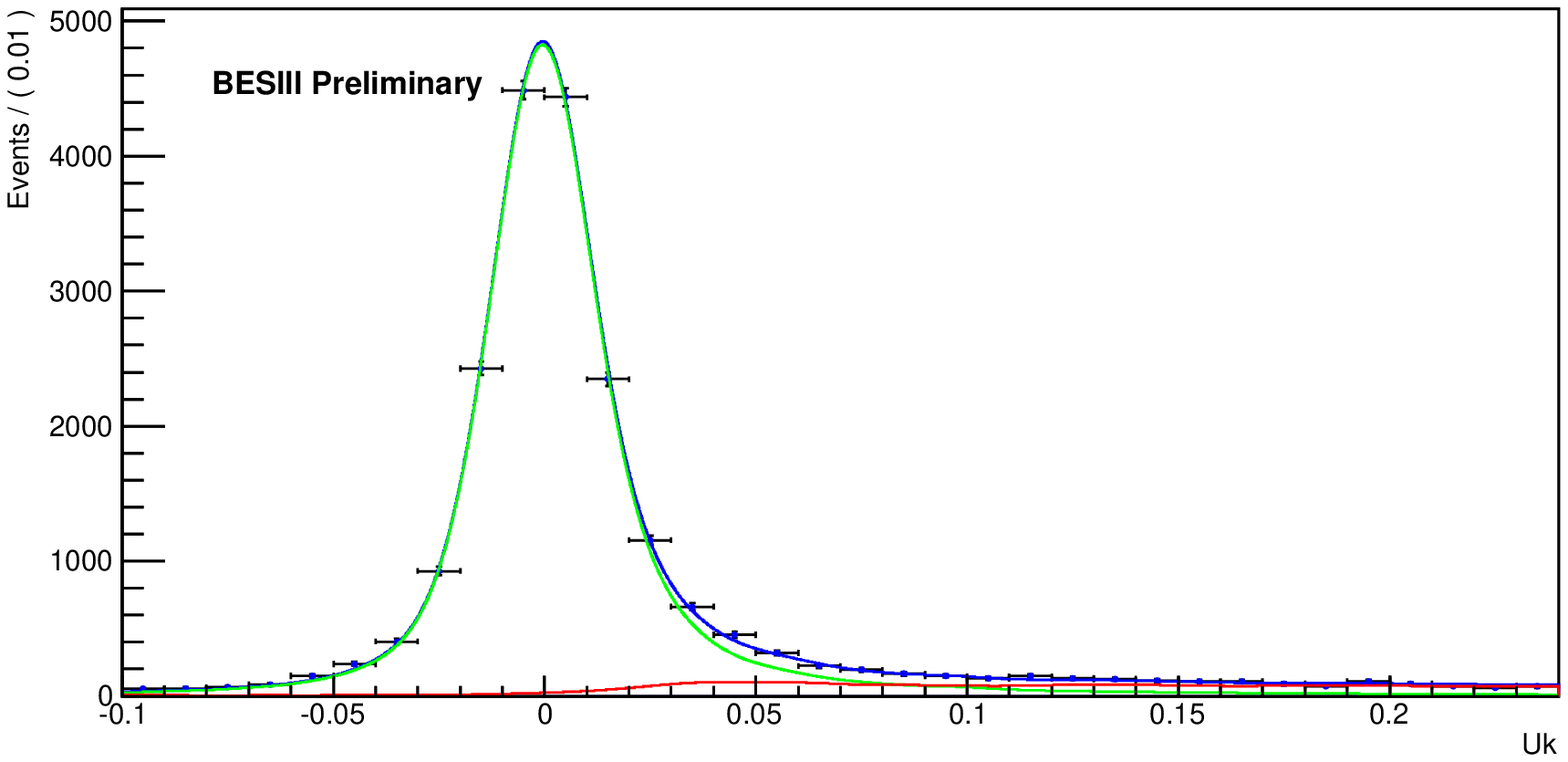}
\includegraphics[height=1.5in,width=0.45\textwidth]{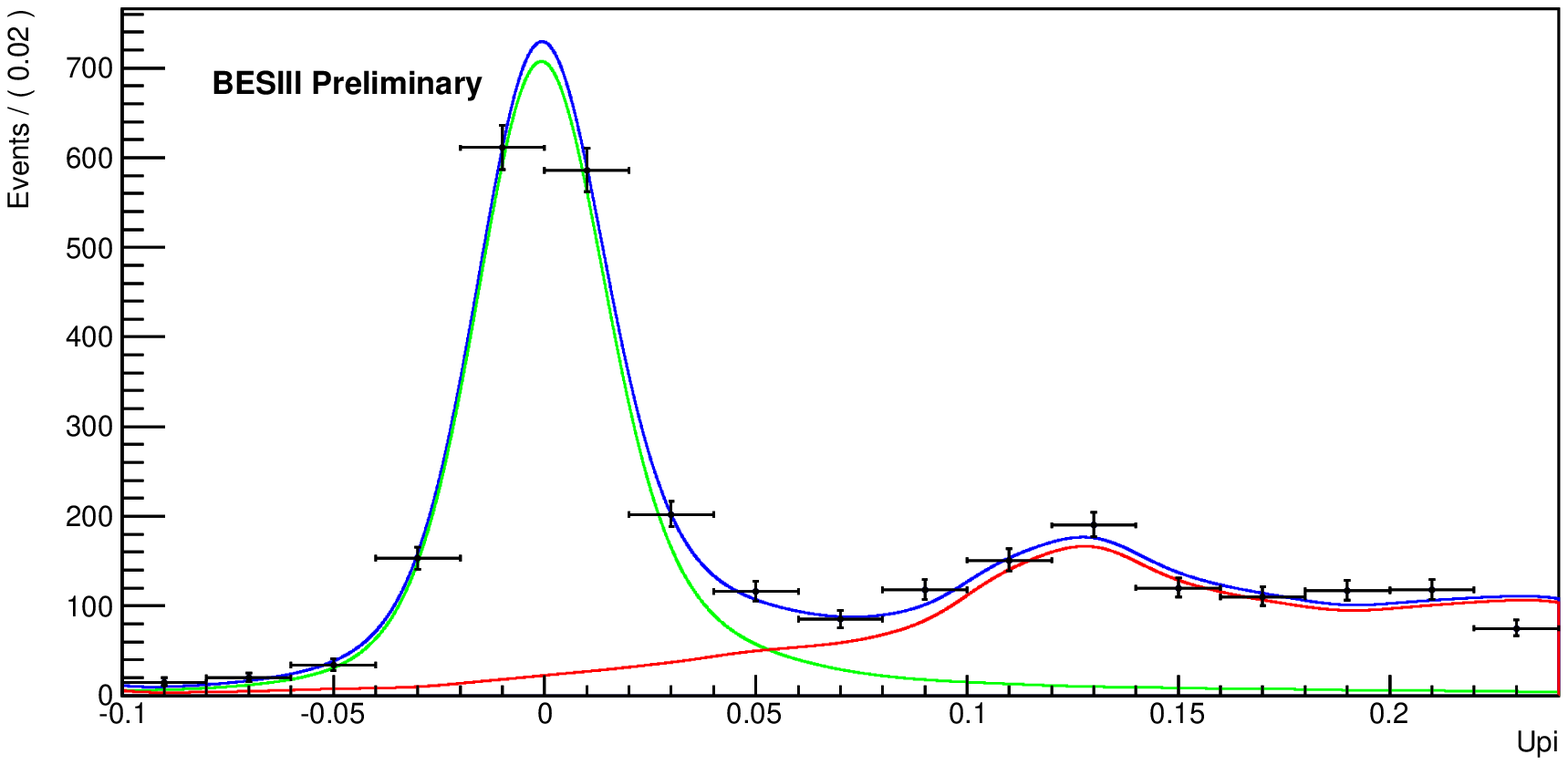}
\caption{ 
$U$ distributions of  $\bar{D}^0\rightarrow K^+ e^- \bar{\nu}$(left) and $\bar{D}^0\rightarrow \pi^+ e^- \bar{\nu}$(right). 
Blue, green, and red curves are the total fit, signal fit, and background fit, 
respectively.
}
\label{fig:u}
\end{figure}

The energy and momentum of the missing neutrino is inferred by using:
\begin{eqnarray}
E_{miss} \;=\; E_{beam}- E_{hadron}- E_{electron} \\ 
\vec{P}_{miss} \;=\; - \vec{P}_{tag} -\vec{P}_{hadron} - \vec{P}_{electron}.
\end{eqnarray}
The number of signal events are obtained by fitting the $U\equiv E_{miss}- P_{miss}$ distributions. Fig.~\ref{fig:u} shows the $U$ distributions and fit projections
for the decay of $\bar{D}^0 \rightarrow K^+ e^- \bar{\nu}$ and $\bar{D}^0 \rightarrow \pi^+ e^- \bar{\nu}$.

\subsubsection{Measurements of Branching Fractions and Partial Decay Rates}

Given the signal yields obtained from fitting $U$ distributions and signal efficiencies obtained from signal Monte Carlo, the branching fractions are obtained through the 
following equation:
\begin{equation}
B_{sig} \;=\;  \frac{N^{obs}_{sig}}{\sum_{\alpha} N^{obs,\alpha}_{tag} \epsilon^{\alpha}_{tag,sig}/\epsilon^{\alpha}_{tag} },
\label{eqn:bf}
\end{equation}
where $N^{obs}_{sig}$ is the total number of signal yields with all tag modes combined, $N^{obs,\alpha}_{tag}$ is the observed tag yields for tag mode $\alpha$,
$\epsilon^{\alpha}_{tag}$ is the tag efficiency for mode $\alpha$, and $\epsilon^{\alpha}_{tag,sig}$ is the combined signal and tag efficiency for mode $\alpha$. 
Preliminary results of branching fractions are listed in Tab.~\ref{tab:bf}.

\begin{table}[htb]
\begin{center}
\begin{tabular}{l|ccc}  
 mode &  Measurement(\%) &  PDG value (\%) & CLEO-c value (\%) \\ \hline
 $\bar{D}^0\rightarrow K^+ e^- \bar{\nu} $ &   3.542$\pm$0.030$\pm$0.067     &     3.55$\pm$0.04     &    3.50$\pm$0.03$\pm$0.04  \\
 $\bar{D}^0\rightarrow \pi^+ e^- \bar{\nu} $ &   0.288$\pm$0.008$\pm$0.005     &    0.289$\pm$0.008     &  0.288$\pm$0.008$\pm$0.003  \\
 \hline
\end{tabular}
\caption{Branching fraction measurements using $\sim923$ pb$^{-1}$ of $\psi(3770)$ data from BESIII.}
\label{tab:bf}
\end{center}
\end{table}

\begin{figure}[htb]
\centering
\includegraphics[height=1.5in,width=0.45\textwidth]{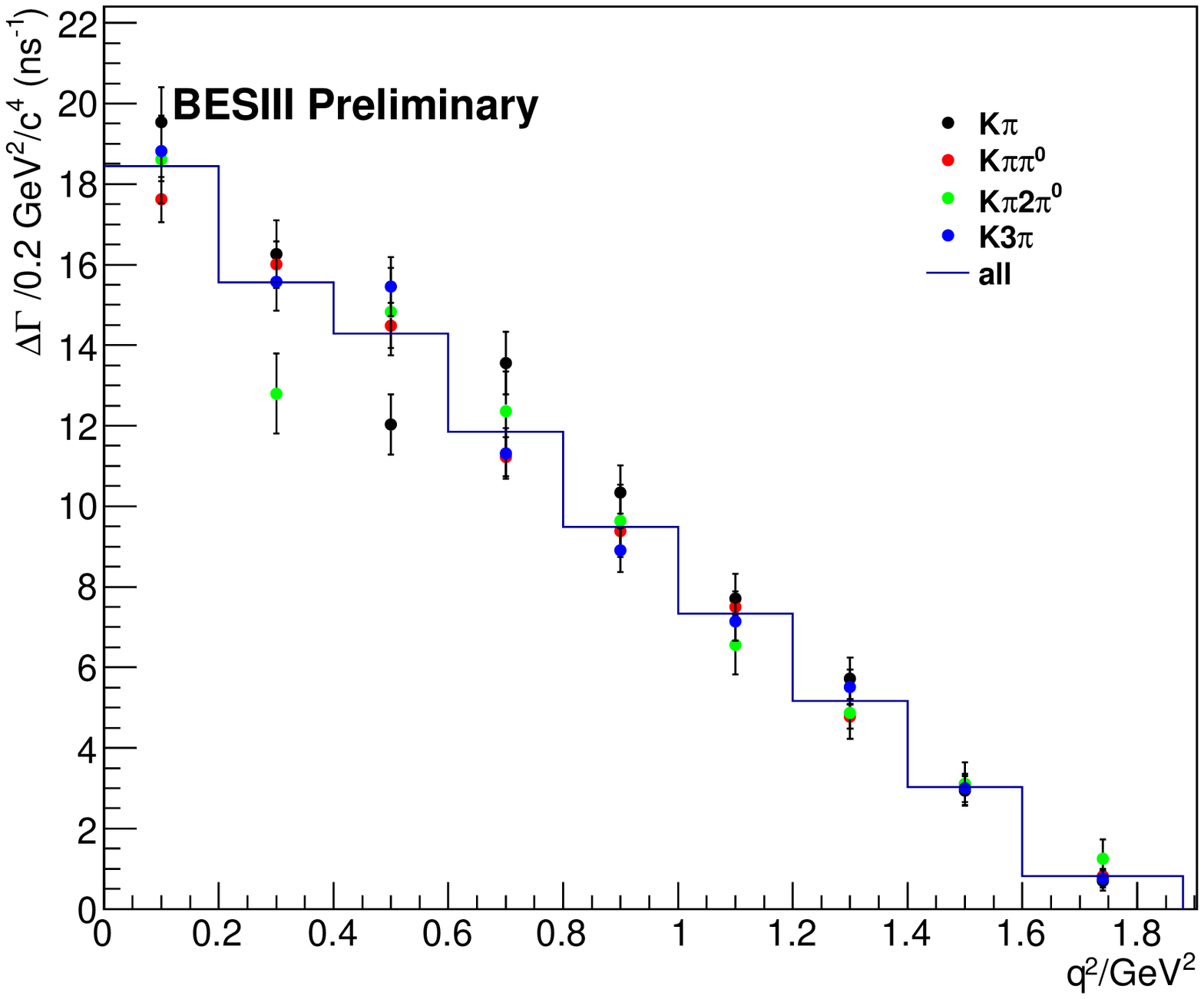}
\includegraphics[height=1.5in,width=0.45\textwidth]{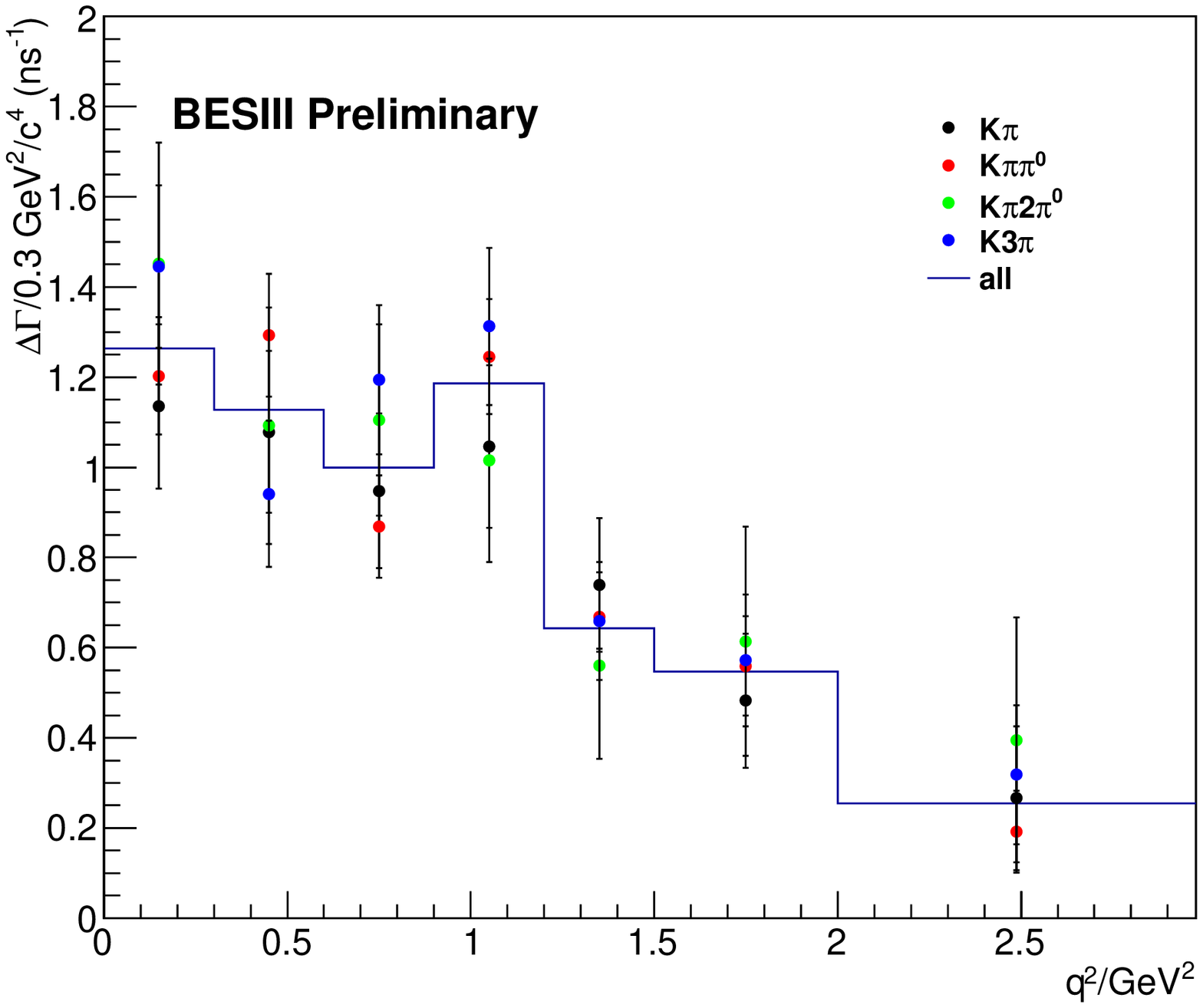}
\caption{ 
Partial decay rates measurement using individual tag modes (points) and all tag modes combined (histogram) for decay of  $\bar{D}^0\rightarrow K^+ e^- \bar{\nu}$ (left) and $\bar{D}^0\rightarrow \pi^+ e^- \bar{\nu}$ (right). 
}
\label{fig:part}
\end{figure}

In order to measure form factor, partial decay rates are measured in different $q^2$ bins, where $q^2$ is the invariant mass squared of the electron-neutrino system.
$ \bar{D}^0\rightarrow K^+ e^- \bar{\nu} $ candidates are divided into nine $q^2$ bins (in GeV$^2$): [0.0,0.2), [0.2,0.4), [0.4,0.6), [0.6,0.8), [0.8,1.0), [1.0,1.2), 
[1.2,1.4), [1.4,1.6), [1.6,$\infty$) , while $\bar{D}^0\rightarrow \pi^+ e^- \bar{\nu} $ are divided into seven $q^2$ bins: [0.0,0.3), [0.3,0.6), [0.6,0.9), [0.9,1.2), [1.2,1.5), [1.5,2.0), 
[2.0,$\infty$). Signal yields in each $q^2$ bin are obtained by fitting $U$ distributions in that $q^2$ range. Using an efficiency matrix vs. $q^2$, obtained
from Monte-Carlo simulation, to correct for smearing, and combining with 
tag yields and tag efficiencies from previous studies, 
the partial decay rates are obtained, as shown in Fig.~\ref{fig:part}.

\begin{figure}[htb]
\centering
\includegraphics[height=1.5in,width=0.45\textwidth]{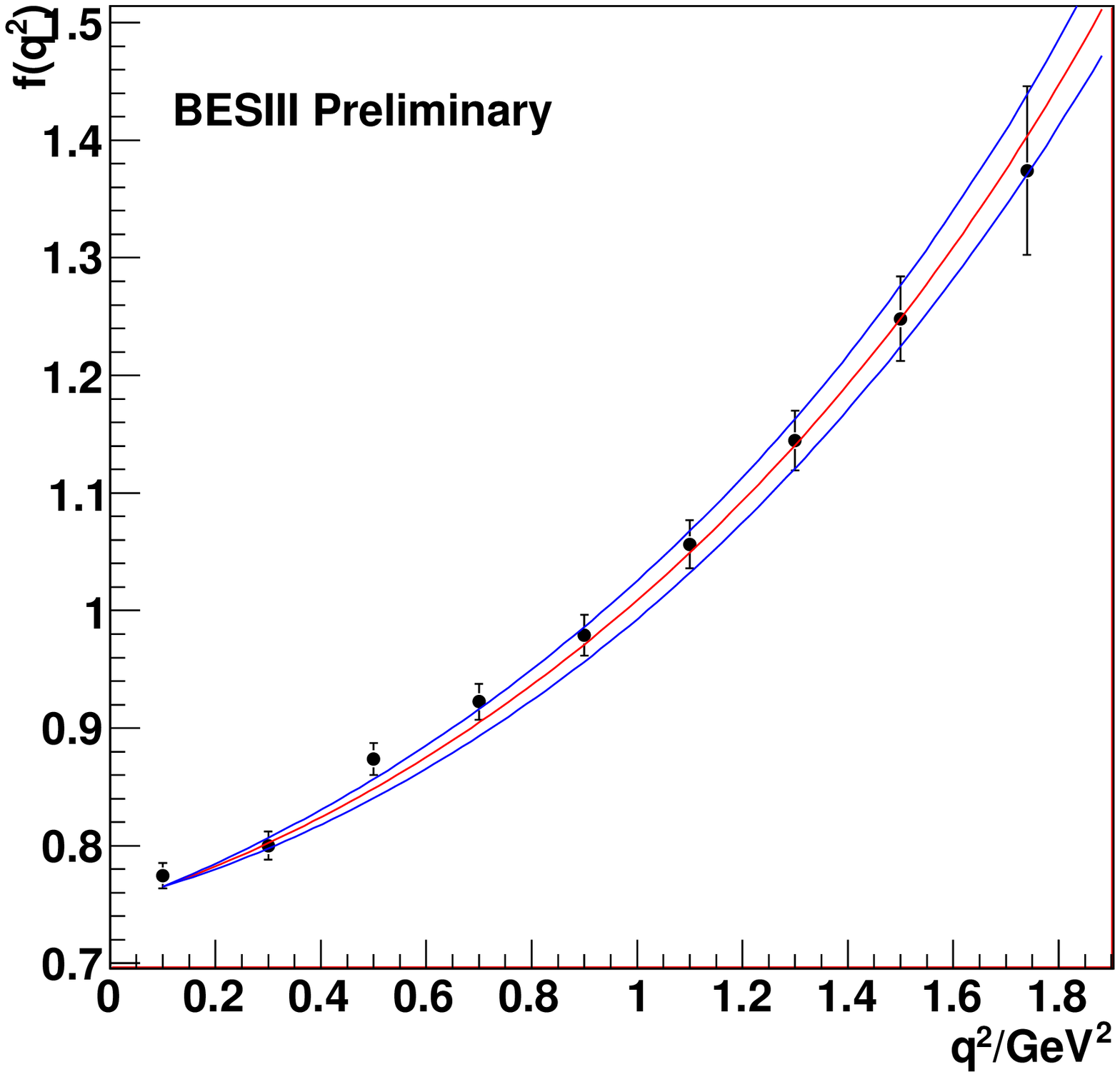}
\includegraphics[height=1.5in,width=0.45\textwidth]{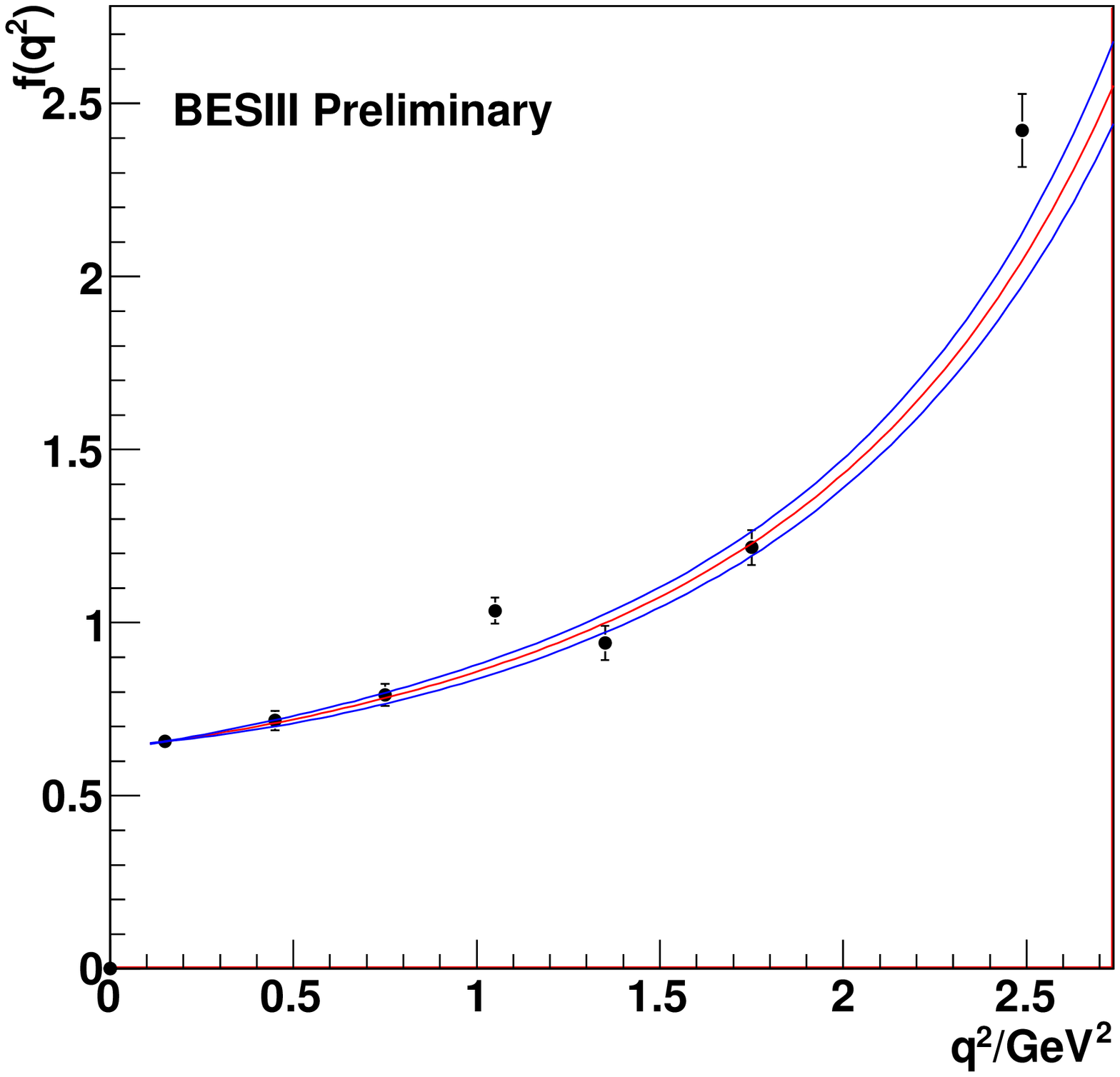}
\caption{ 
 $f_+(q^2)$ distributions for the decay of  $\bar{D}^0\rightarrow K^+ e^- \bar{\nu}$ (left) and $\bar{D}^0\rightarrow \pi^+ e^- \bar{\nu}$ (right). 
Points are measured from data, red curves are the theoretical predictions, and blue curves represents variations within one statistical standard sigma.
}
\label{fig:fq2}
\end{figure}

With measured partial decay rates, we can also obtain the $f_+(q^2)$ at the center 
of each $q^2$ bin $i$ using: 
\begin{eqnarray}
f_+(q^2_i) \;=\; \frac{1}{|V_{cd(s)}|}\sqrt{\frac{\Delta\Gamma_i}{\Delta q^2_i}\frac{24\pi^3}{G^2_F p^3_i}},
\end{eqnarray}
where $|V_{cd(s)}|$ can be taken from the PDG value, and $p^3_i$ is the effective value averaged over 
$q^2_i$, calculated from:
\begin{equation}
p^3_i \;=\; \frac{\int p^3|f_+(q^2)|^2 dq^2}{\int|f_+(q^2)|^2 dq^2},
\end{equation}
where the three-parameter series parameterization for $f_+(q^2)$ is used from fitting data later.
The $f_+(q^2)$ distributions are shown in Fig.~\ref{fig:fq2} with theoretical curves overlaid~\cite{fermifq2}.

\subsubsection{Measurements of the Form Factors}

Three different parameterizations of the form factor $f_{+}(q^2)$ are considered.

The first parameterization, known as the simple pole model, is dominated by a single
pole~\cite{ff-singlepole}:
\begin{eqnarray}
 f_+(q^2) & = & \frac{f_+(0)}{1- q^2/m^2_{pole} },
\end{eqnarray}
where the values of $m_{pole}$ is predicted to be $M_{D^*_{(s)}}$.  
Note that, however, that $m_{pole}$ is floated in the fit in addition to $f_+(0)$  
to improve fit quality.  

The second parameterization is known as the modified pole model~\cite{ff-singlepole}:
\begin{eqnarray}
 f_+(q^2) & = & \frac{f_+(0)}{
                    \left( 1- \frac{q^2}{m^2_{pole}} \right)
                    \left(1- \alpha\,\frac{q^2}{m^2_{pole}} \right)
                             },
\end{eqnarray}
where $m_{pole}$ is generally fixed to the $D^*_{(s)}$ mass.  
Both $f_+(0)$ and $\alpha$ are free parameters.

The third parameterization is known as the series expansion~\cite{ff-series}. 
In the series expansion model, the branch cut in the complex $q^2$ plane 
is mapped onto a unit circle $|z|<1$, where
\begin{eqnarray}
  z(q^2,t_0) \,\,=\,\,  
     \frac{ \sqrt{t_+ - q^2} - \sqrt{t_+ - t_0} }
          { \sqrt{t_+ - q^2} + \sqrt{t_+ - t_0} },
\end{eqnarray}
where $t_{\pm} = (m_D \pm m_P)^2$, 
and $t_0$ is any real number smaller than $t_+$.  
This transformation expands the form factor about $q^2=t_0$ as:
\begin{eqnarray}
  f_+(q^2) \,\,=\,\,  
   \frac{1}{ P(q^2) \, \phi(q^2,t_0)} \,\,
   \sum^{\infty}_{k=0} \,\, a_k(t_0) \,\, \left[ z(q^2,t_0) \right]^k,
\end{eqnarray}
where $a_l$ are real coefficients, 
$P(q^2) = z(q^2, m^2_{D^*})$ for kaon final states, 
$P(q^2) = 1$ for pion final states, 
and $\phi(q^2,t_0)$ is any function that is analytic outside a cut 
in the $q^2$ plane that lies long the x-axis from $t_+$ to $\infty$.  
The standard choices are $t_0 \,=\, t_+(1-(1-t_-/t_+)^{1/2})$ 
and, for $\phi$: 
\begin{eqnarray}
 \phi(q^2,t_0) \,\,=\,\, 
   \sqrt{ \frac{\pi m^2_c}{3}} \,\, 
         \left( \frac{z(q^2,0)}{-q^2} \right)^{5/2} \,
         \left( \frac{z(q^2,t_0)}{t_0-q^2} \right)^{-1/2} \, 
      \left( \frac{z(q^2,t_-)}{t_--q^2} \right)^{-3/4} \,
       \frac{t_+-q^2}{(t_+-t_0)^{1/4}}
\end{eqnarray}
Notice that the term $\frac{z(q^2,0)}{-q^2}$ has a limit 
of $1/(4t_+)$ as $q^2$ goes to zero.

With the above parameterization, the data is fitted by minimizing this 
$\chi^2$ function:
\begin{eqnarray}
\chi^2 &  \,\,=\,\, &  \sum^{n}_{i,j=1} \,\,
       \left( \Delta\Gamma_i - \Delta G_i \right) \,\, C^{-1}_{ij} \,\, 
       \left (\Delta\Gamma_j - \Delta G_j \right),
\end{eqnarray}
where $\Delta\Gamma_i$ is the measured partial decay rate in $q^2$ bin $i$, 
$\Delta G_i$ is the predicted partial decay rate in $q^2$ bin $i$, 
and $C$ is the covariance matrix of the measured partial decay rates.
The fitted results are listed in Tab.~\ref{tab:formfit}.

\begin{table}[htb]
\begin{center}
\begin{tabular}{l|c|c|c}
\hline  
Simple Pole &  $f_+(0)|V_{cd(s)}|$  &  $m_{pole}$  &  \\
 $\bar{D}^0\rightarrow K^+ e^- \nu$ &   0.729$\pm$0.005$\pm$0.007     &    1.943$\pm$0.025$\pm$0.003    &     \\
 $\bar{D}^0\rightarrow \pi^+ e^- \nu$ & 0.142$\pm$0.003$\pm$0.001     &    1.876$\pm$0.023$\pm$0.004    &     \\
\hline
Modified Pole &  $f_+(0)|V_{cd(s)}|$  &  $\alpha $  &  \\
 $\bar{D}^0\rightarrow K^+ e^- \nu$ &   0.725$\pm$0.006$\pm$0.007     &    0.265$\pm$0.045$\pm$0.006    &     \\
 $\bar{D}^0\rightarrow \pi^+ e^- \nu$ & 0.140$\pm$0.003$\pm$0.002     &    0.315$\pm$0.071$\pm$0.012    &     \\
\hline
2 par. series &  $f_+(0)|V_{cd(s)}|$  &  $r_1 $  &  \\
 $\bar{D}^0\rightarrow K^+ e^- \nu$ &   0.726$\pm$0.006$\pm$0.007     &    -2.034$\pm$0.196$\pm$0.022    &     \\
 $\bar{D}^0\rightarrow \pi^+ e^- \nu$ & 0.140$\pm$0.004$\pm$0.002     &    -2.117$\pm$0.163$\pm$0.027    &     \\
\hline
3 par. series &  $f_+(0)|V_{cd(s)}|$  &  $r_1 $  &  $r_2$ \\
 $\bar{D}^0\rightarrow K^+ e^- \nu$ &   0.729$\pm$0.008$\pm$0.007     &    -2.179$\pm$0.355$\pm$0.053    & 4.539$\pm$8.927$\pm$1.103     \\
 $\bar{D}^0\rightarrow \pi^+ e^- \nu$ & 0.144$\pm$0.005$\pm$0.002     &    -2.728$\pm$0.482$\pm$0.076    &  4.194$\pm$3.122$\pm$0.448   \\
\hline
\end{tabular}
\caption{
Fitter parameters from form factor measurements for $\bar{D}^0\rightarrow K^+ e^- \nu$ and $\bar{D}^0\rightarrow \pi^+ e^- \nu$.
}
\label{tab:formfit}
\end{center}
\end{table}

\subsection{Measurement of $D^+\rightarrow K^- \pi^+ e^+ \nu_e$}
\label{sec:vector}

The $D^+\rightarrow K^- \pi^+ e^+ \nu$ decay includes contributions from 
different $K\pi$ resonances, including the dominant $\bar{K}^*(892)^0$ 
piece, as well as a non-resonant amplitude.  
The measurement of the $\bar{K}^*(892)^0$ (lowest vector meson) 
contribution can be used to study hadronic transition form factors. Using 347.5 fb$^{-1}$ of data recorded at the $\Upsilon(4S)$ 
the BaBar collaboration has analyzed this decay channel~\cite{paper-barbar}.  
In addition to measuring the form factors, a comprehensive analysis of 
many other items is also performed.  
These include the properties of the $\bar{K}^*(892)^0$ meson, 
the $S$-wave contribution to the decay, 
the variations of the $K\pi$ $S$-wave phase versus the $K\pi$ mass, 
and a search for radially excited $P$-wave and $D$-wave $K\pi$ resonances.  

The formalism of the differential decay partial width is taken from Ref.~\cite{paper-barbar2}, which necessarily includes five variables:
\begin{itemize}
\item $m^2$, the squared mass of the kaon-pion system;
\item $q^2$, the squared mass of the electron-neutrino system;
\item cos($\theta_{K}$), where $\theta_K$ is the angle between the kaon
three-momentum in the kaon rest frame and the
line of flight of the kaon-pion system in the $D$ rest frame;
\item cos($\theta_e$), where $\theta_e$ is the angle between the electron three-momentum in the electron-neutrino rest frame and
the line of flight of the electron-neutrino system in the $D$ rest frame;
\item $\chi$, the angle between the normals to the planes
defined in the $D$ rest frame by the $K\pi$ pair and the $e\nu$ pair.
\end{itemize}

\subsubsection{ Event Selection}
To reconstruct $D^+$ decays, all the charged and neutral particles are boosted to the center-of-mass system and the event thrust axis is determined.  
A plane perpendicular to this axis defines two hemispheres, and the candidate hemisphere consists of a positron, a charged kaon, and a charged 
pion. A vertex fit is performed and events with probability larger than 10$^{-7}$ are kept.  To estimate the neutrino momentum,
the $K^- \pi^+ e^+ \nu_e$ system is constrained to the $D^+$ mass with estimates of the $D^+$ direction and neutrino energy obtained
from all tracks and unmatched showers measured in the event.  
The neutrino energy is evaluated by subtracting from the hemisphere energy 
the energy of reconstructed particles contained in that hemisphere.  
To further reject $B\bar{B}$ background and continuum background 
(arising mainly from charm particles), 
Fisher discriminants are created from relevant variables.  This discriminant retains 40\% of signal events and rejects 94\% of the remaining background.
About 244$\times 10^3$ signal events are selected with a final signal-to-background ratio of 2.3\,.

\subsubsection{\boldmath Measurement of Form Factors and $\bar{K}^*(892)^0$ Properties}

Form factors $F_{1,2,3}$ can be expanded into partial waves to show explicit dependence on $\theta_K$. Considering only
$S$, $P$ and $D$ waves, this gives:
\begin{eqnarray}
F_1 \;=\;  F_{10} + F_{11}cos\theta_K +  F_{12} \frac{3cos^2\theta_K -1}{2}   \\ \nonumber
F_2 \;=\;  \frac{1}{\sqrt{2}} F_{21} +  \sqrt{\frac{3}{2}} F_{22} cos\theta_K \\ 
F_3 \;=\;  \frac{1}{\sqrt{2}} F_{31} +  \sqrt{\frac{3}{2}} F_{32} cos\theta_K   \nonumber
\end{eqnarray}
where $F_{10}$ characterizes the $S$-wave contribution, and $F_{i1}$ and $F_{i2}$ correspond to the $P-$ and $D$-waves, respectively.
It is also possible to relate these form factors with the helicity form factors $H_{0,\pm}$, which can be in turn related 
to the two axial-vector form factors $A_{1,2}(q^2)$ and the vector form factor $V(q^2)$. For the $q^2$ dependence, the simple pole
parameterization is used:
\begin{eqnarray}
V(q^2)   \;=\;  \frac{V(0)}{1-q^2/m^2_V}  , \\  \nonumber
A_1(q^2) \;=\;  \frac{A_1(0)}{1-q^2/m^2_A}  , \\  
A_2(q^2) \;=\;  \frac{A_2(0)}{1-q^2/m^2_A}  , \nonumber
\end{eqnarray}
where $m_V$ and $m_A$ are expected to be close to $m_{D^*_s} = 2.1$ GeV and $m_{D_{s1}}=2.5$ GeV, respectively. In the analysis, ratios
of these form factors, evaluated at $q^2=0$, $r_{V}= V(0)/A_1(0)$ and $r_{2}= A_2(0)/A_1(0)$ are measured by studying the variation of 
partial decay rate versus kinematic variables; $m_A$ is allowed to fit while $m_V$ is fixed. For the mass dependence, in case of the 
$K^*(892)$, a Breit-Wigner distribution is used.  

Tab.~\ref{tab:barbar1} shows the fit results considering three different models: 
\begin{itemize}
\item  $S + \bar{K}^*(892)^0$, a signal made of the $\bar{K}^*(892)^0$ and $S$-wave components.
\item  $S + \bar{K}^*(892)^0+ \bar{K}^*(1410)^0 $, a signal made of the $\bar{K}^*(892)^0$,  $\bar{K}^*(1410)^0$,  and $S$-wave components.
\item  $S + \bar{K}^*(892)^0+ \bar{K}^*(1410)^0 + D $, a signal made of the $\bar{K}^*(892)^0$,  $\bar{K}^*(1410)^0$,  $S$- and $D$-wave components.
\end{itemize}  
The fractions of signal components in different models are measured and shown in Tab.~\ref{tab:barbar2}. The second model is considered as the
nominal fit to data, with systematics error also listed in the table.

\begin{table}[htb]
\begin{center}
\begin{tabular}{l|ccc}  
Variable & $S + \bar{K}^*(892)^0$   &  $S + \bar{K}^*(892)^0 $  & $S + \bar{K}^*(892)^0 $  \\ 
         &                           &  $\bar{K}^*(1410)^0 $    & $\bar{K}^*(1410)^0 + D $  \\ 
\hline

 $m_{K^*(892)}$ (MeV) &   894.77$\pm$0.08     &   895.4$\pm$0.2$\pm$0.2   & 895.27$\pm$0.21   \\
 $\Gamma^0_{K^*(892)}$( MeV) &   45.78$\pm$0.23     &   46.5$\pm$0.3$\pm$0.2   & 46.38$\pm$0.26   \\
 $r_{BW}$ (GeV$^{-1}$) &   3.71$\pm$0.22     &   2.1$\pm$0.5$\pm$0.5   & 2.31$\pm$0.20   \\
 $m_A$ (GeV)          &   2.65$\pm$0.10     &  2.63$\pm$0.10$\pm$0.13   & 2.58$\pm$0.09    \\
 $r_V$                &   1.458$\pm$0.016     &  1.463$\pm$0.017$\pm$0.031   & 1.471$\pm$0.016    \\
 $r_2$                &   0.804$\pm$0.020     &  0.801$\pm$0.020$\pm$0.020   & 0.786$\pm$0.020    \\

\hline
\end{tabular}
\caption{ Values of form factor and $\bar{K}^*(892)^0$ parameters with different models.}
\label{tab:barbar1}
\end{center}
\end{table}

\begin{table}[htb]
\begin{center}
\begin{tabular}{l|ccc}  
Component & $S + \bar{K}^*(892)^0$   &  $S + \bar{K}^*(892)^0 $  & $S + \bar{K}^*(892)^0 $  \\ 
         &                           &  $\bar{K}^*(1410)^0 $    & $\bar{K}^*(1410)^0 + D $  \\ 
\hline
  $S$-wave              &  5.62$\pm$0.14$\pm$0.13   &  5.79$\pm$0.16$\pm$0.15    &  5.69$\pm$0.16$\pm$0.15   \\
  $P$-wave              &  94.38   &  94.21    &  94.12   \\
 $\bar{K}^*(892)^0$   &  94.38   &  94.11$\pm$0.74$\pm$0.75    &  94.41$\pm$0.15$\pm$0.20  \\
 $\bar{K}^*(1410)^0$  &  0       &  0.33$\pm$0.13$\pm$0.19     &  0.16$\pm$0.08$\pm$0.14  \\
  $D$-wave              &  0       &  0                          &  0.19$\pm$0.09$\pm$0.09  \\
\hline
\end{tabular}
\caption{ Fractions (in percent) of signal components with different models.}
\label{tab:barbar2}
\end{center}
\end{table}

\subsubsection{\boldmath Measurement of $S$-wave Phase}

\begin{figure}[htb]
\centering
\includegraphics[height=2.0in,width=0.6\textwidth]{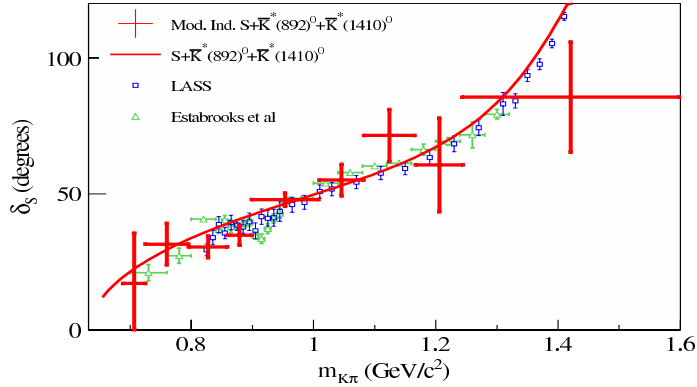}
\caption{ 
Points (crosses) give the $S$-wave phase variation using the nominal model.  
The phase is assumed to be constant within each mass bin, and
parameters of the $\bar{K}^*(892)^0$ are fixed to values from the nominal fit to data.  The solid line corresponds to the parameterized $S$-wave
phase variation in the nominal fit.  LASS data are shown in blue.  
}
\label{fig:barbarphase}
\end{figure}

Using the nominal signal model containing $S$-wave, $\bar{K}^*(892)^0$ and $\bar{K}^*(1410)^0$ components, the phase in
the mass-dependent $S$-wave amplitude is measured and compared to LASS measurements. The S-wave phase is assumed to be
a constant within each of the considered $K\pi$ mass intervals. 
The measurement of phase variation is shown in Fig.~\ref{fig:barbarphase}.

\subsection{\boldmath Measurement of  $D^0/D^+ \rightarrow \rho e \nu$ and $D^+ \rightarrow \omega e \nu$}

Using 818 pb$^{-1}$ of data taken at the $\psi(3770)$, the CLEO-c collaboration has measured branching fraction and form factor for 
the decays of  $D^0/D^+ \rightarrow \rho e \nu$ and the branching fraction 
for $D^+ \rightarrow \omega e \nu$~\cite{paper-cleoc}. 
The precision of the branching fractions is improved, and the form factor result of the $D^0/D^+ \rightarrow \rho e \nu$ 
is the first measurement on the Cabibbo-suppressed 
pseudo-scalar meson to vector meson transition in semi-leptonic $D$ decay. 

\subsubsection{Event Selection}
\begin{figure}[htb]
\centering
\includegraphics[height=2.5in,width=0.8\textwidth]{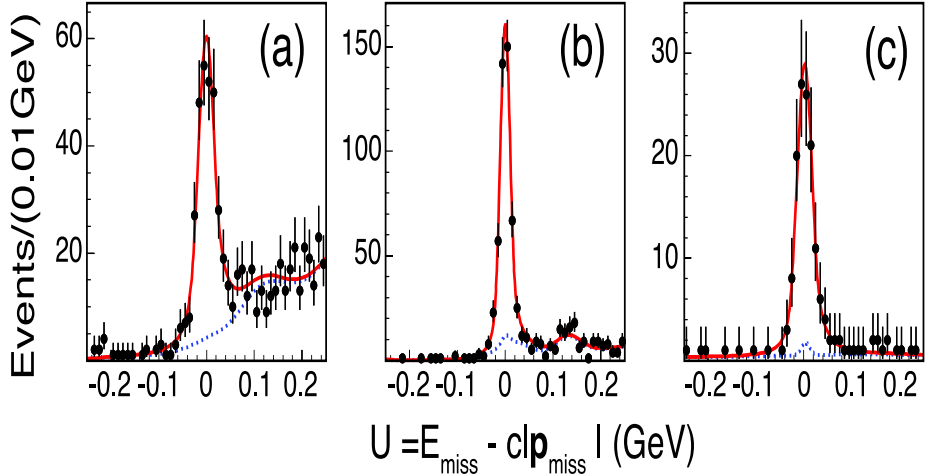}
\caption{ 
$U$ distributions of (a) $D^0 \rightarrow \rho^- e^+ \nu_e$, (b) $D^+\rightarrow \rho^0 e^+ \nu_e$  and (c) $D^+ \rightarrow \omega e^+ \nu$.
}
\label{fig:umiss-rho}
\end{figure}

The double-tag technique is used in this analysis by reconstructing a $D$ tag in the following hadronic final states: 
$K^+\pi^-$, $K^+\pi^-\pi^0$, and $K^+\pi^-\pi^-\pi^+$ for neutral tags, and $K^0_s\pi^-$, $K^+\pi^-\pi^-$,  $K^0_s\pi^-\pi^0$, $K^+\pi^-\pi^-\pi^0$, $K^0_s \pi^-\pi^-\pi^+$,  
and $K^-K^+\pi^-$ for charged tags. In case of multiple candidates in the same tag mode, the candidate with minimum $\Delta E$ is chosen. Once a tag is
identified, certain $\Delta E$ and $m_{BC}$ cuts are required. The unused tracks and showers are then searched for a candidate $e^+$ along with a $\rho^-(\pi^-\pi^0)$, $\rho^0(\pi^+\pi^-)$,
or $\omega(\pi^+\pi^-\pi^0)$. The $\rho$ candidate is required to have invariant mass within 150 MeV from the nominal PDG mass. The combined tag and semi-leptonic 
candidates must account for all reconstructed tracks and unmatched showers in the event. To remove multiple candidates in each semi-leptonic mode, one combination is chosen
per tag mode per tag charge, based on the proximity of the invariant masses of the $\rho^0$, $\rho^+⁺$ or $\omega$ candidates to their PDG masses.
 The number of semi-leptonic decays are obtained by fitting the $U\equiv E_{miss}- |\vec{p}_{miss}|$
distributions, where $E_{miss}$ and $\vec{p}_{miss}$ are energy and momentum of the missing neutrino, and they can be inferred from all other measured particles.
The $U$ distributions are shown in Fig.~\ref{fig:umiss-rho}.

\subsubsection{Measurements of Branching Fraction}
To measure the absolute branching fraction, Monte-Carlo samples are used to determine tag and signal efficiencies. Together with the tag and signal yields,
the same equation as Eq.~\ref{eqn:bf} can be used to give the branching fraction. The measured results are listed in Tab.~\ref{tab:bfrho}.

\begin{table}[htb]
\begin{center}
\begin{tabular}{l|ccccc}  
Decay mode & $\epsilon$(\%)   &  $N_{sig}$  & $B_{SL}$  & $B_{SL}$(ISGW2)   & $B_{SL}$(FK)  \\
 
\hline
   $D^0 \rightarrow \rho^- e^+ \nu_e$     &  26.03$\pm$0.02   &  304.6$\pm$20.9    &  1.77$\pm$0.12$\pm$0.10  &  1.0  & 2.0   \\
   $D^+ \rightarrow \rho^0 e^+ \nu_e$     &  42.84$\pm$0.03   &  447.4$\pm$24.5    &  2.17$\pm$0.12$^{+0.12}_{-0.22}$  &  1.3  & 2.5   \\
   $D^+ \rightarrow \omega e^+ \nu_e$     &  14.67$\pm$0.03   &  128.5$\pm$12.6    &  1.82$\pm$0.18$\pm$0.07  &  1.3  & 2.5   \\

\hline
\end{tabular}
\caption{ Branching fractions for  $D^0 \rightarrow \rho^- e^+ \nu_e$, $D^+\rightarrow \rho^0 e^+ \nu_e$  and $D^+ \rightarrow \omega e^+ \nu$, from the CLEO-c analysis and two model predictions: ISGW2\cite{rhopaper1} and FK\cite{rhopaper2}. The uncertainties for signal efficiency $\epsilon $ and signal yields
$N_{sig}$ are statistical only. The efficiency includes the $\rho$ and $\omega$ branching fractions from the PDG.
}
\label{tab:bfrho}
\end{center}
\end{table}

\subsubsection{Measurements of Form Factor}
 
\begin{figure}[htb]
\centering
\includegraphics[height=2.5in,width=0.7\textwidth]{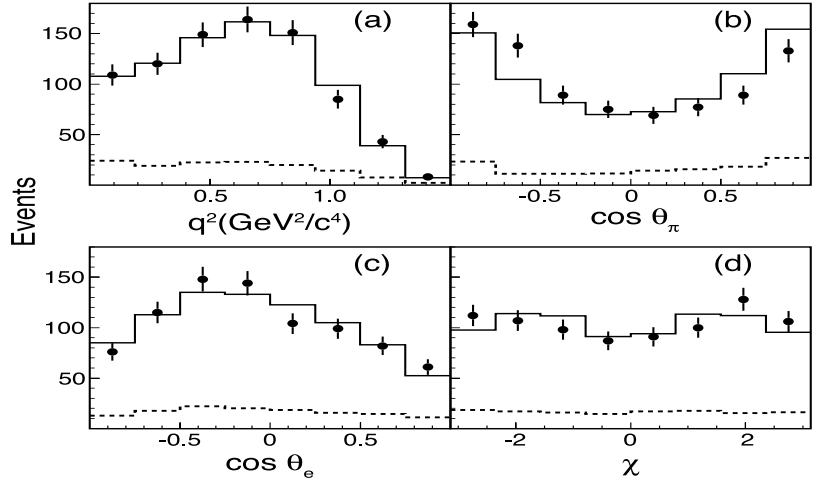}
\caption{ 
Projections of the combined $\rho^-$ and $\rho^0$ data (points with statistics error bar) and the fit (solid histogram). The
dashed lines show the background distributions.
}
\label{fig:rhofit}
\end{figure}

A form factor analysis is performed for $D^0/D^+ \rightarrow \rho^-/\rho^0 e^+ \nu_e$ decays.
The mechanism of this decay is similar to the $D^+\rightarrow K\pi e \nu$ discussed in Sec.~\ref{sec:vector}, except only $P$-wave is considered in this case.
Three dominant form factors, two axial and one vector, $A_1$, $A_2$, and $V$, 
are used to describe the hadronic current. A simple pole model is assumed with
the pole mass fixed as $M_{D^*(1^-)} = 2.01$ GeV and $M_{D^*(1^+)} = 2.42$ GeV for the vector and axial form factors, respectively. 
A four-dimensional maximum likelihood fit~\cite{rhoLL} is performed in the space of $q^2$. $cos\theta_{\pi}$, $cos\theta_e$ and $\chi$, and a simultaneous fit
is made to the isospin-conjugate modes $D^0 \rightarrow \rho^0 e^+ \nu_e$ and $D^+ \rightarrow \rho^- e^+ \nu_e$.  The ratios of form factors evaluated at 
$q^2=0$, $r_V = \frac{V(0)}{A_1(0)}$ and $r_2 = \frac{A_2(0)}{A_1(0)}$ are obtained:  $r_V= 1.48 \pm 0.15\pm0.05$ and $r_2 = 0.83\pm0.11\pm0.04$.
Using $|V_{cd}|=0.2252\pm0.0007$, 
$\tau_{D^0} = (410.1\pm 1.5)\times10^{-15}$ s, 
and $\tau_{D^+} = (1040\pm 7)\times 10^{-15}$ s,  
from PDG 2010, form factor ratios and branching fraction results 
are combined to obtain : $A_1(0)=0.56\pm0.01^{+0.02}_{-0.03}$, $A_2(0) = 0.47\pm0.06\pm0.04$, and $V(0)=0.84\pm0.09^{+0.05}_{-0.06}$. 
The fit projection is shown in Fig.~\ref{fig:rhofit}, where the difference between data and the fit projection for $cos\theta_{\pi}$ may be due to $S$-wave
interference.

\subsection{Search for the Decay of $D^+_s\rightarrow \omega e^+ \nu$ }
 
Using data collected at a center-of-mass energy $\sqrt{s} = 4170$ MeV, CLEO-c 
searched for the decay of $D^+_s\rightarrow \omega e^+ \nu$, which can probe 
the four-quark content of the $D^+_s$~\cite{search}.  The integrated luminosity of the data used in this analysis is 586 pb$^{-1}$, or
0.6$\times 10^6$ $D^+_sD^-_s$ inclusive decays. Recent work by Gronau and Rosner~\cite{search2} concludes that any value of the branching fraction
for $D^+_s\rightarrow \omega e^+ \nu$ exceeding 2$\times 10^{-4}$ is unlikely to be from $\omega-\phi$ mixing.

\subsubsection{Event Selection}
About 95\% of the $e^+e^- \rightarrow D^+_s X$ events are formed from the following exclusive reactions~\cite{dsdecay}:
\begin{equation}
e^+e^- \rightarrow D^+_s D^{*-}_s,\    \ e^+e^- \rightarrow D^-_s D^{*+}_s,
\end{equation}
with equal amounts of them produced, and about 95\% of the $D^*_s$ decays through $D_s \gamma$,

To select signal events, the double tag technique is used. Eight $D_s$ tag modes are reconstructed: 
$K^0_sK^-$, $K^+K^-\pi^-$, $K^{*-}\bar{K}^{*0}$, $\pi^+\pi^-\pi^-$, $\eta\pi^-$, $\eta\rho^-$, 
$\pi^-\eta^{\prime}(\eta\pi^+\pi^-)$, $\pi^-\eta^{\prime}(\rho\gamma)$. The four-momentum of a tag candidate is 
defined by $(E_{tag}, \vec{p}_{tag})$, and the recoil mass $M_{rec}$ is defined as 
\begin{equation}
M_{rec} \;=\; \sqrt{ (E_b-E_{tag})^2 - (\vec{p}_b - \vec{p}_{tag})^2 },
\end{equation}
where $(E_b,\vec{p}_b)$ is the four-momentum of the colliding beams. The $M_{rec}$ distribution peaks at zero only for events
where the photon is associated with signal side, but even if the photon is associated with the tag side, $M_{rec}$ is 
kinematically constrained such that $|M_{rec}- M_{D^*_s}|< 55$ MeV. So only events passing with this wide cut are retained.
The distribution of invariant mass of the tag $M_{tag}$ is fitted with a double Gaussian function, and the weighted
resolution $\sigma$ is used to define signal regions. All tag modes are required to to be within 2.5$\sigma$ from 
the peak position except for the $\eta\rho$ mode where it is selected with 2$\sigma$. A second kinematic constraint is used, on the $MM^{*2}$ defined as
\begin{equation}
MM^{*2} \;=\; (E_b - E_{tag} - E_{\gamma})^2 - (\vec{p}_b -\vec{p}_{tag} - \vec{p}_{\gamma})^2,
\end{equation} 
where $MM^{*2}$ is expected to peak at $M^2_{D_s}$ (independent of 
whether the $\gamma$ is from the tag or signal side).  
The selection criteria on $MM^{*2}$ is determined from a 
two-dimensional binned likelihood fit the $(MM^{*2},M_{tag})$ space.  

The signal is then selected by requiring one positron candidate with charge opposite to the tag charge, two charged pions of
opposite charges, and no extra good tracks. A yet-unused good $\pi^0$ candidate is also required. 
There can be multiple $\pi^0$ candidates; if so, the one with lowest
$\chi^2 \;=\; [(M_{\gamma\gamma}-M_{\pi^0})/\sigma_{\gamma\gamma}]^2$ is chosen, 
where $M_{\gamma\gamma}$ is the photon-photon mass, 
and $\sigma_{\gamma\gamma}$ is the calculated resolution.  
The positron, charged pions, and $\pi^0$ are added together
to get a four-vector $(E_s,\vec{p}_s)$. With information from beams and tag side, the missing-mass squared of the neutrino is obtained: 
\begin{equation}
MM^2  \;=\; (E_b - E_{tag}- E_{\gamma}- E_s)^2 - (\vec{p}_b - \vec{p}_{tag}- \vec{p}_{\gamma} - \vec{p}_s)^2.
\end{equation}

\subsubsection{Final Fit and Results}

\begin{figure}[htb]
\centering
\includegraphics[height=2.5in,width=0.8\textwidth]{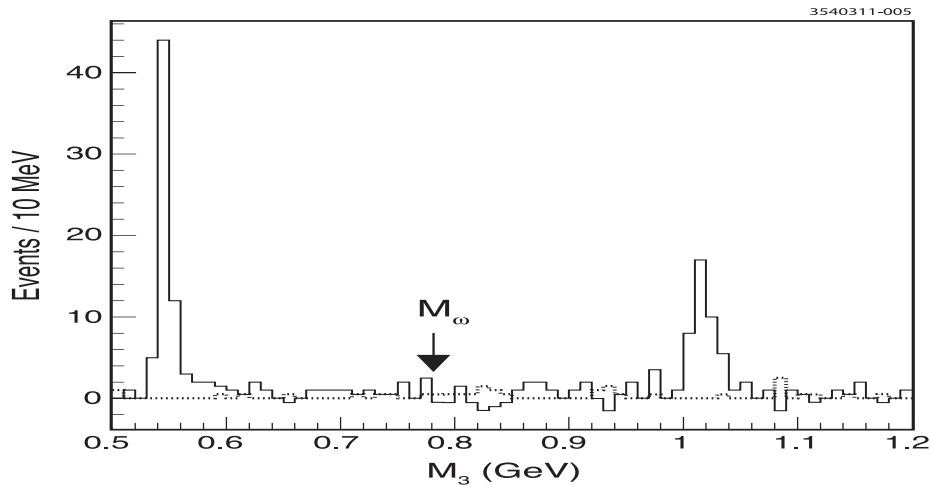}
\caption{ 
$M_3$ ($M(3\pi$) distribution. Solid line: signal selection, after $M_{tag}$ sideband subtraction. Dotted line: $M_{tag}$ sideband contribution.
}
\label{fig:m3}
\end{figure}

To extract the signal yields, the mass of the $\pi^+\pi^-\pi^0$ system, $M_3$, 
is used with a cut on $MM^2$ as $-0.05 < MM^2 < 0.05$ GeV$^2$.  
The $M_3$ distribution is shown in Fig.~\ref{fig:m3},
where one can see peaks for the $\eta$ and $\phi$, but no clear signal peak at 
the $\omega$. An upper limit on the 
branching fraction of $B(D^+_s \rightarrow \omega e^+ \nu)<0.20\%$ is obtained at 90\% confidence level.

\section{Summary}

Experimental measurements of semi-leptonic $D$ decays have been studied 
very successful during the past few years, with contributions 
from experiments such as FOCUS, Belle, BaBar and CLEO-c. 
During this presentation, recent results on semi-leptonic $D$ decays 
have been discussed, with topic covering several topical aspects, 
i.e., pseudo-scalar to pseudo-scalar modes,
pseudo-scalar to vector modes, and rare modes. 

Since the start of running in 2008, the newest of these experiments, 
BESIII, has taken about 2.9 fb$^{-1}$ of data at $\psi(3770)$.  
With peak luminosity reaching 
more than $6 \times 10^{32}$ (60\% of the designed luminosity),
BESIII is poised to take more data at $\psi(3770)$ and in the higher $D_s$ 
energy region.  
Using part of the data, BESIII has presented preliminary results of
the $D^0\rightarrow K/\pi e \nu$ decays. 
Results from the full dataset and other modes are coming in the near future.

\Acknowledgements

The author thanks Charm 2012 committee for the invitation to the conference and
their great organization.

\end{document}